\documentclass{aa}

\usepackage[varg]{txfonts}
\usepackage{graphicx}

\bibpunct{(}{)}{;}{a}{}{,}

\begin{document}

\title{ $^{16}$O/$^{18}$O ratio in water in the coma of comet 67P/Churyumov-Gerasimenko measured with the Rosetta/ROSINA double-focusing mass spectrometer}

\author{Isaac R.H.G. Schroeder I\inst{1}
        \and Kathrin Altwegg\inst{1,2}
        \and Hans Balsiger\inst{1}
        \and Jean-Jacques Berthelier\inst{3}
        \and Johan De Keyser\inst{4}
        \and Bj{\"o}rn~Fiethe\inst{5}
        \and Stephen A. Fuselier\inst{6,7}
        \and S{\'e}bastien Gasc\inst{1}
        \and Tamas I. Gombosi\inst{8}
        \and Martin Rubin\inst{1}
        \and Thierry S{\'e}mon\inst{1}
        \and Chia-Yu~Tzou\inst{1}
        \and Susanne F. Wampfler\inst{2}
        \and Peter Wurz\inst{1,2}}

\institute{Physikalisches Institut, University of Bern, Sidlerstrasse 5, CH-3012 Bern, Switzerland
        \and Center for Space and Habitability, University of Bern, Gesellschaftsstrasse 6, CH-3012 Bern, Switzerland
        \and LATMOS, 4 Avenue de Neptune, F-94100 Saint-Maur, France
        \and Royal Belgian Institute for Space Aeronomy (BIRA-IASB), Ringlaan 3, B-1180, Brussels, Belgium
        \and Institute of Computer and Network Engineering (IDA), TU Braunschweig, Hans-Sommer-Stra{\ss}e 66, D-38106 Braunschweig, Germany
        \and Space Science Division, Southwest Research Institute, 6220 Culebra Road, San Antonio, TX 78228, USA
        \and University of Texas at San Antonio, San Antonio, TX
        \and Department of Atmospheric, Oceanic and Space Sciences, University of Michigan, 2455 Hayward, Ann Arbor, MI 48109, USA}

\date{Received 12 July 2018 / Accepted 30 August 2018}

\abstract{The European Space Agency spacecraft Rosetta accompanied the Jupiter-family comet 67P/Churyumov-Gerasimenko for over two years along its trajectory through the inner solar system. Between 2014 and 2016, it performed almost continuous in situ \mbox{measurements} of the comet’s gaseous atmosphere in close proximity to its nucleus. In this study, the $^{16}$O/$^{18}$O ratio of H$_2$O in the coma of 67P/Churyumov-Gerasimenko, as measured by the ROSINA DFMS mass spectrometer on board Rosetta, was determined from the ratio of  H${_2}^{16}$O / H${_2}^{18}$O and $^{16}$OH / $^{18}$OH. The value of 445 $\pm$ 35 represents an $\sim$11\% enrichment of $^{18}$O compared with the \mbox{terrestrial} ratio of 498.7 $\pm$ 0.1. This cometary value is consistent with the comet containing primordial water, in accordance with leading \mbox{self-shielding} models. These models predict primordial water to be between 5\% to 20\% enriched in heavier oxygen isotopes compared to terrestrial water.}


\titlerunning{$^{16}$O/$^{18}$O ratio in water in the coma of comet 67P}
\authorrunning{I. Schroeder et al.}
\maketitle

\section{Introduction}

Comets are widely considered to contain some of the most \mbox{pristine} material in the solar system \citep{wyckoff1991}. The degree of isotopic fractionation, that is, the enrichment or depletion of an isotope in a molecule relative to its initial abundance, observed in a comet is sensitive to the environmental conditions at the time of the comet’s formation \citep{haessig2017}. Therefore, measurements of isotopic abundances in cometary ices reveal important information regarding the composition, density, and temperature of the early solar system. These measurements also indicate the amount of radiation that was present during the accretion of solid bodies, when the molecules were being formed during the chemical evolution of the presolar cloud to the protosolar nebula and protoplanetary disk. They are therefore vital to understanding and reconstructing the history and origins of material in the solar system, which was one of the major scientific goals of the Rosetta mission \citep{glassmeier2007}. Oxygen is of particular interest to us because large heterogeneities in its relative isotopic abundance in meteoritic samples have frustrated efforts to determine the primordial composition of the solar system, and a lack of correlation with presolar components suggests that
isotope-selective chemistry occurred within the protosolar nebula \citep{mckeegan2011}. \\

The European Space Agency (ESA) spacecraft Rosetta accompanied the Jupiter-family comet (JFC) designated 67P/Churyumov-Gerasimenko (hereafter 67P) for a period of two years. Between August 2014 and September 2016, the spacecraft studied its coma and nucleus in great detail during its orbit around the Sun from its approach at around 3.5 AU to its perihelion passage and then out to 3.5 AU. The Rosetta Orbiter Spectrometer for Ion and Neutral Analysis (ROSINA) mass spectrometers on board, designed to measure isotopic abundances \citep{balsiger2007}, continuously analyzed the volatile species in the cometary coma for almost the entirety of this duration. \\

With its high mass resolution, dynamic range, and \mbox{sensitivity}, ROSINA was able to detect rare species such as HD$^{18}$O alongside their most abundant isotopologs \citep{haessig2013}, and measure isotopic ratios in water such as D/H and $^{16}$O/$^{18}$O independently. It was already able to measure the deuterium-to-hydrogen (D/H) ratio in cometary water very early on in its mission, finding a D/H ratio of more than three times the terrestrial value. This vital result revealed much about the comet’s origin, the water formation temperature, and the conditions under which the early solar system formed \citep{altwegg2017}. It also showed that JFCs have a wide range of D/H ratios and was thus an important measurement for the discussion of the origins of terrestrial oceans. \\

Oxygen is the most abundant element not only in the solid phases that formed early in the development of the solar system \citep{yurimoto2004}, but also in rocky materials in general, because its cosmic abundances and the affinity between O and Si are high. The $^{16}$O/$^{18}$O ratio of CO$_2$ in the coma of comet 67P was previously measured by \citet{haessig2017}  with Rosetta’s ROSINA instrument package Double Focusing Mass Spectrometer (DFMS) and found to be 494 $\pm$ 8, which is consistent within 1$\sigma$ uncertainty with the terrestrial value of 498.7 $\pm$ 0.1 calculated by \citet{baertschi1976}. In contrast, the solar wind has a $^{16}$O/$^{18}$O ratio of 530 $\pm$ 2 \citep{mckeegan2011}. A more detailed list of measurements for other comets is provided in Fig. \ref{fig4}. \\

Here, we report on the results of direct in situ measurements of the $^{16}$O/$^{18}$O ratio in H$_2$O from the coma of 67P, performed with the Rosetta / ROSINA DFMS.

\section{Instrumentation and method}

The ROSINA DFMS is a double-focusing mass spectrometer with a high mass resolution of m/$\Delta$m $\sim$ 3000 at 1\% peak height \citep{balsiger2007}. Neutral gas entering the DFMS is ionized via electron impact ionization with an electron energy of 45\,eV, which causes a certain percentage of parent molecules to split into charged fragments. (Fragmentation patterns are species-specific, unique to each spectrometer, and dependent on the electron energy.) The ions and fragments then pass through an electrostatic analyzer and permanent magnet and are filtered by their mass-to-charge ratio before reaching the detectors. \\

The primary detector, MCP / LEDA, is a position-sensitive imaging detector comprised of two micro-channel plates (MCPs) in a chevron configuration. When ions impinge on the MCPs, they release a cascade of secondary electrons, thereby amplifying their signal, which is then detected by two \mbox{independent} rows (Row A and B) of 512 anodes on a linear electron detector array (LEDA). Row B serves as a redundancy to Row A. The voltage applied between the front and back of the MCP can be adjusted to vary the gain (degree of amplification) of the MCP detector. Measurements are not all obtained at the same detector gain: there are 16 predefined voltage settings \mbox{referred} to as gain steps (GS1 to GS16), and the DFMS measures by scanning over a range of masses one at a time, automatically selecting the gain step for each mass that maximizes the signal without causing saturation. \\

The gain corresponding to a certain gain step changes over time because the detector ages. This time-dependency necessitated dividing the mission into intervals and using time-interpolation between tables of different gain correction factors that were separately derived for each interval. \\

An additional flat-field correction was also required to account for the uneven degradation of the 512 LEDA anodes (pixels) with use because those in the center were used more frequently and were consequently more degraded. This was referred to as the pixel gain to distinguish it from the (overall) gain. Special modes of the DFMS were dedicated to the measurement of pixel gain; they measured water at a fixed gain step by slowly shifting the center of the peak from one end of the array to the other so as to compare the amplitude of the signal detected by each of the pixels. Campaigns to measure the pixel gain were conducted at regular intervals, and linear time-interpolation was applied to derive the correction factors at other times. \\

As a result of spacecraft outgassing, Rosetta had a neutral gaseous background (primarily water with traces of organic \mbox{material}, hydrazine from thruster exhaust and fluorine from vacuum grease). This background had a permanent particle density of $\sim$10$^6$ cm\textsuperscript{–3} in the immediate vicinity of the spacecraft, even prior to its rendezvous with the comet \citep{altwegg2014}. Even after ten years of traveling through the vacuum of space while en route to 67P, the gaseous background from Rosetta could still be measured and characterized with ROSINA \citep{schlaeppi2010}, demonstrating its ability to analyze even trace amounts of gases. The isotopic composition of the water outgassed from the Rosetta spacecraft itself was consistent, as expected, with terrestrial values \citep{haessig2013}, and it did not vary with the time of degassing, indicating negligible \mbox{isotope} fractionation. Exploiting this fact, we were able to use the \mbox{fragmentation} pattern of outgassed water, from \mbox{pre-encounter} measurements acquired during a sniff test on 19$^{}$~June~2014, as a reference for correcting subsequent \mbox{measurements}. \\

The DFMS is not equally sensitive to all masses. It has different relative sensitivities for each mass, which apparently changed as it aged. Abrupt changes, such as one observed on 3$^{}$ June 2015, can also be caused by damage. By comparing isotopic ratios measured during the sniff test with their expected (terrestrial) values, corrections for changes in relative sensitivity were derived. These were then used to correct the measured fragmentation pattern of water. \\

An accurate determination of how water fragments in the DFMS is important for the calculation of the gain. The ratio of $^{18}$OH / H${_2}^{18}$O from the sniff test was 0.33 $\pm$ 0.04. We chose to compare the amount of H${_2}^{18}$O detected with that of the $^{18}$OH produced by its fragmentation because (unlike H${_2}^{16}$O and $^{16}$OH) they were often both measured on the same gain step (GS16), and their ratio in such instances would not be affected even if the gain factors were incorrect. However, this fragmentation pattern should be the same for $^{16}$OH / H${_2}^{16}$O, in addition to staying constant throughout the mission. Thus, any differences between later measurements and this value reflect changes in the gain and were accordingly used to derive corrections to it. Further details regarding the data analysis and corrections we applied may be found in the appendix. \\

By incorporating all these corrections to the gain, pixel gain, and relative sensitivity in our model of the instrument aging, we calculated the $^{16}$O/$^{18}$O ratio of H$_2$O by taking the average ratio of H${_2}^{16}$O / H${_2}^{18}$O and $^{16}$OH / $^{18}$OH over both rows of the DFMS MCP / LEDA detector between 1$^{}$ October 2014 and 5$^{}$~September 2016. This period spans almost the entirety of the duration in which Rosetta was in close proximity to comet 67P. \\

\vspace{1cm}

\begin{figure*}
\centering
        \includegraphics[width=17cm]{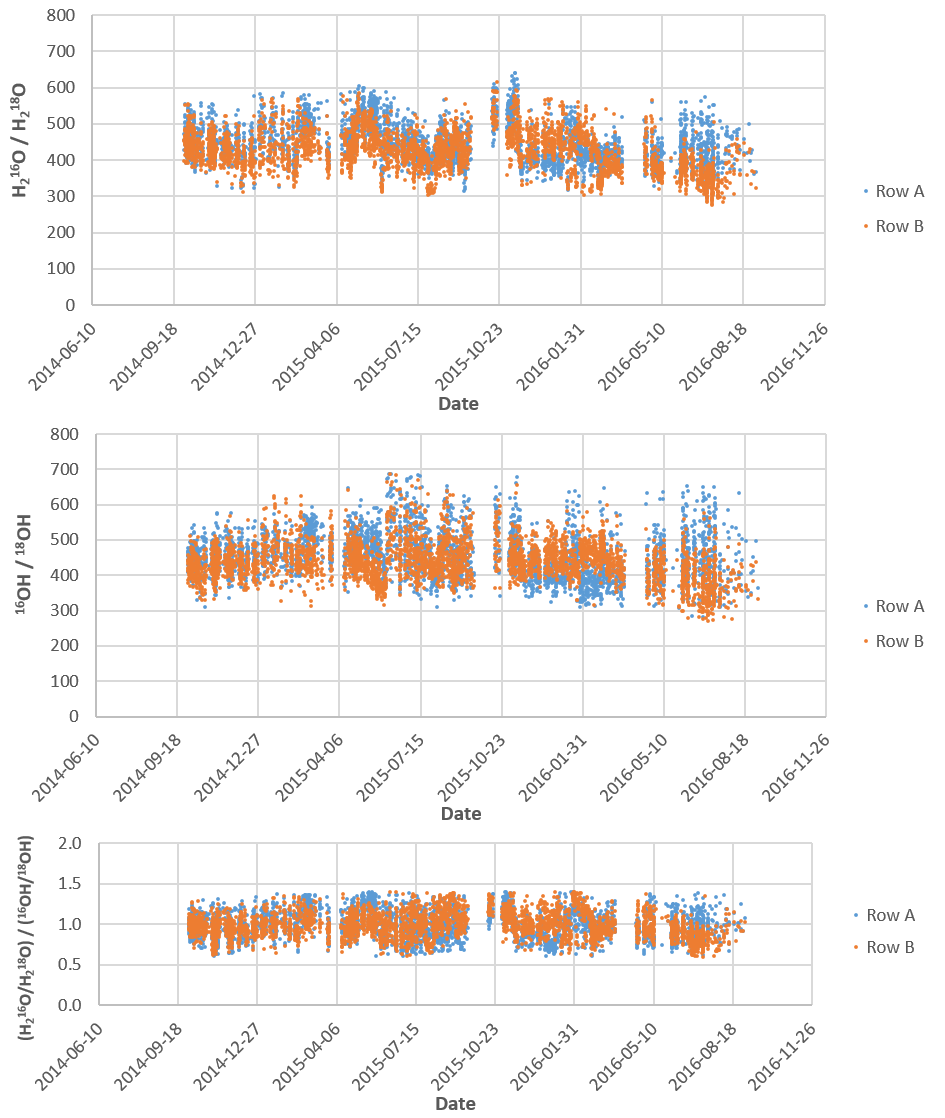}
        \caption{H${_2}^{16}$O / H${_2}^{18}$O ratio (top), $^{16}$OH / $^{18}$OH ratio (middle), and \textbf{$\frac{H{_2}^{16}O / H{_2}^{18}O}{^{16}OH / ^{18}OH}$} ratio (bottom) over the course of the mission. See \textbf{Fig. \ref{fig8}} in the appendix for an idea of the typical error bars on a single isotopic ratio measurement.}
        \label{fig1}
\end{figure*}

\section{Results}

We found an average $^{16}$O/$^{18}$O ratio of H$_2$O in the coma of comet 67P of 445 $\pm$ 35. This result was based on 3820 \mbox{measurements} of H${_2}^{16}$O / H${_2}^{18}$O and $^{16}$OH / $^{18}$OH, which were in close agreement with each other. The measurements were made from 1$^{}$ October 2014 to 5$^{}$ September 2016 with both Rows (A and B) of the MCP / LEDA. That both ratios were consistent with each other despite having been measured on \mbox{different} gain steps shows that the detector aging model is accurate. \\

The $^{16}$O/$^{17}$O ratio, however, could not be estimated, \mbox{unfortunately}, as the signal from H${_2}^{17}$O was too low in addition to being buried in the shoulder of the much larger HDO peak. \\

The 3820 individual values of the isotopic ratio measured over the course of the main part of the Rosetta mission are shown in \textbf{Fig. \ref{fig1}} for the H${_2}^{16}$O / H${_2}^{18}$O and $^{16}$OH / $^{18}$OH ratios. Despite the large spread in the data, no obvious change over time could be discerned from the isotopic ratios. Here, we also made the assumption that the detected OH is predominantly a product of H$_2$O fragmentation. This assumption is justified on the grounds that contributions from other possible parent molecules (e.g. alcohols) are negligible, as they are several orders of magnitude less abundant than H$_2$O and their fragmentation only produces OH at low to intermediate levels (2\% for methanol, 12\% for ethanol), which we established in our calibration experiments. \\

\begin{figure}
        \resizebox{\hsize}{!}{\includegraphics{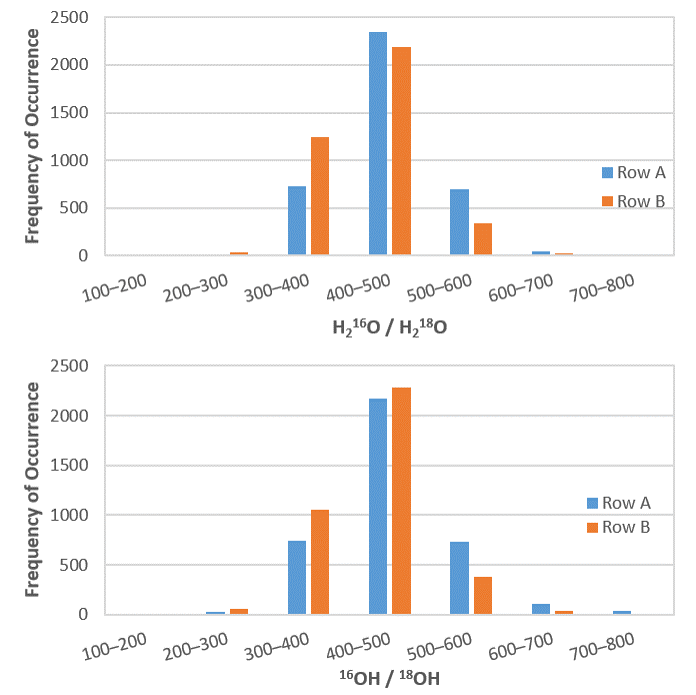}}
        \caption{Histograms of individual measurements (top: H${_2}^{16}$O / H${_2}^{18}$O , bottom: $^{16}$OH / $^{18}$OH).}
        \label{fig3}
\end{figure}

The individual measurements from Fig. \ref{fig1} are also shown in \textbf{Fig. \ref{fig3}} in the form of histograms depicting the range of measured values. \\

Many factors contributed to the large spread seen in the distribution of data points over the course of the mission and the consequently large uncertainties in the isotopic ratios. As the DFMS measures each mass separately, factors such as \mbox{spacecraft} motion and measurement time affected the data analysis. Also affecting the analysis were instrument effects arising from difficulties in correcting for the pixel-dependent degradation of the MCP / LEDA detector (the pixel gain correction) and changes in the gain of the detector over time. The uncertainties in the gain (6\%) and pixel gain (5\%) were the dominant sources of error, while the statistical uncertainty is roughly two to three orders of magnitude smaller because of averaging over a large number of measurements. \\

\newpage

\section{Discussion}

The value of 445 $\pm$ 35 found for the $^{16}$O/$^{18}$O ratio of cometary H$_2$O from the coma of 67P represents an enrichment of \, approximately 11\%  of $^{18}$O compared with the terrestrial value of 498.7 $\pm$ 0.1 measured by \citet{baertschi1976}. However, the two measurements are also statistically compatible within a 1.5\,$\sigma$ uncertainty. The present value differs from an earlier result, 556~$\pm$~62, reported by \citet{altwegg2014} because of the recent development of a more sophisticated detector aging model. \\

In contrast, the $^{16}$O/$^{18}$O ratio from CO$_2$ measurements of the coma of 67P, as previously performed by \citet{haessig2017}, was 494 $\pm$ 8, which is consistent with the terrestrial value within the uncertainties. The solar wind measurement by \citet{mckeegan2011}, on the other hand, had a $^{16}$O/$^{18}$O ratio of 530 $\pm$ 2. \\

\begin{figure*}
\centering
	\includegraphics[width=17cm]{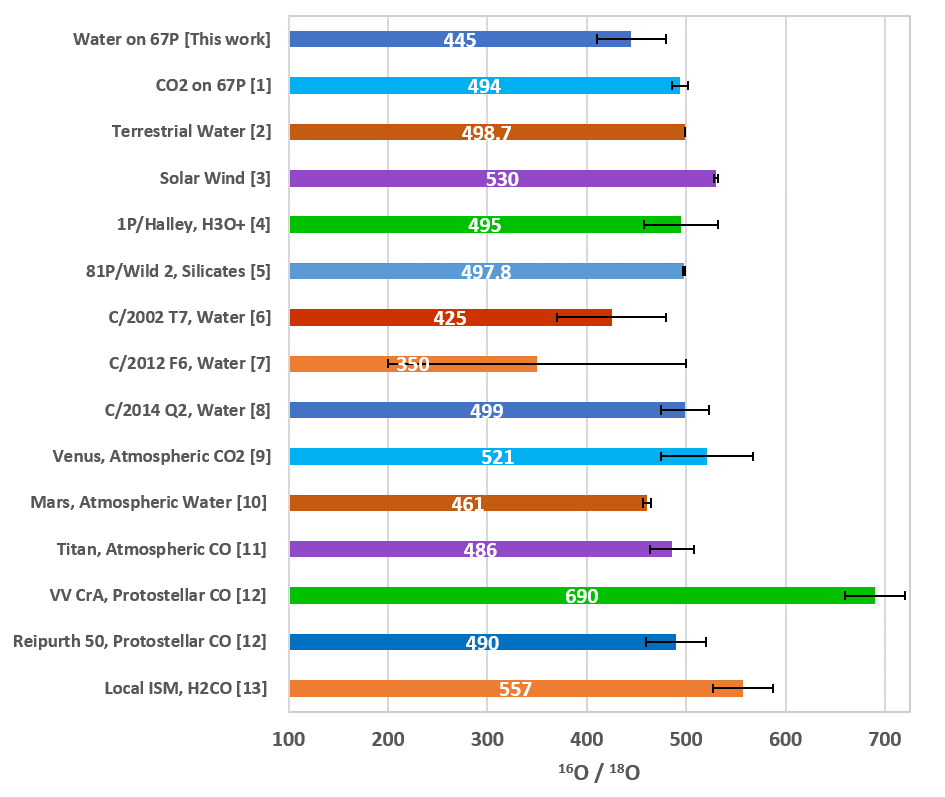}
	\caption{Comparison of $^{16}$O/$^{18}$O ratios from various sources.}
	\label{fig4}
	\tablebib{(1) \citet{haessig2017}; (2) \citet{baertschi1976}; (3) \citet{mckeegan2011}; (4) \citet{eberhardt1995,balsiger1995}; (5) \citet{ogliore2015}; (6) \citet{hutsemekers2008}; (7) \citet{decock2014}; (8) \citet{biver2016}; (9) \citet{iwagami2015}; (10) \citet{webster2013}; (11) \citet{serigano2016}; (12) \citet{smith2009}; (13) \citet{wilson1999}.}
\end{figure*}

A comparison of the $^{16}$O/$^{18}$O ratio of H$_2$O from the coma of comet 67P with the $^{16}$O/$^{18}$O ratios from several other sources and results from preceding publications \citep{bockeleemorvan2015} is provided in \textbf{Fig. \ref{fig4}}. That the isotopic \mbox{fractionation} of CO$_2$ should differ from that of water is unsurprising, since CO$_2$ freezes out at 81 K, a lower temperature than water (160 K) but higher than CO (29 K), in the solar nebula \citep{marboeuf2014,yurimoto2004} and is also chemically derived from CO, which the self-shielding phenomenon discussed below fractionates differently than water. \\

According to leading self-shielding models \citep{lyons2005,young2007,lee2008}, primordial water is predicted to be enriched in $^{18}$O by 5\% to 20\% compared to terrestrial water, whereas the solar wind is expected to be depleted in $^{18}$O by $\sim$5\%  \citep{sakamoto2007,yurimoto2004}. \\

The considerable isotopic fractionation observed for oxygen and carbon in molecular clouds is thought to be the result of self-shielding in the ultraviolet photodissociation of CO \citep{bally1982,vandishoeck1988}. The same effect is expected to have occurred during the T-Tauri stage in the evolution of our Sun \citep{clayton2002}, where the proto-sun provided a strong source of ultraviolet radiation and the gas in the disk was comprised primarily of H$_2$, CO and N$_2$. \\

Self-shielding of CO in the solar nebula involves the isotope-selective photodissociation of CO, which occurs at far-ultraviolet (FUV) wavelengths between 91.2\,nm and 110\,nm \citep{lyons2005,warin1996}. CO can transition (prior to dissociation) to a bound excited state with a lifetime long enough to exhibit vibrational and rotational structure. The resulting absorption spectrum has many narrow absorption lines that are shifted when the molecular mass is altered as a result of isotopic substitution. Additionally, the absorption spectra of the various CO isotopologs do not overlap significantly \citep{lyons2005}. Thus, when a cloud is irradiated by an ultraviolet continuum, the wavelengths corresponding to the most abundant isotopolog, $^{12}$C$^{16}$O, are more rapidly attenuated \citep{clayton2002} by the surface layer of the cloud than those for the less abundant $^{12}$C$^{18}$O. The latter thus penetrate deeper into the cloud, enabling the dissociation of $^{12}$C$^{18}$O to continue even deep in the cloud interior. The dissociation of $^{12}$C$^{16}$O in the interior is meanwhile suppressed as a result of the lack of UV photons with its requisite wavelengths. This produces a zone of $^{18}$O-enriched atomic oxygen (CO dissociates into C and O) and leaves the remaining undissociated CO correspondingly $^{18}$O-depleted. We were unfortunately unable to test this prediction with direct measurements of $^{12}$C$^{18}$O as the resolving power of the DFMS was insufficient to distinguish it from the more abundant $^{14}$N$^{16}$O \citep{rubin2017}. However, the protostar VV CrA in Fig.~\ref{fig4} does indeed conform to this expectation, though the protostar Reipurth 50 does not. The reason for this seeming discrepancy is that the protoplanetary disk was probed in the case of VV~CrA, whereas it was the protostellar envelope that was being probed by the observed CO absorption lines for Reipurth~50 \citep{smith2009}. \\

The dominant oxygen-bearing species of ice, H$_2$O \citep{yurimoto2004,langer2000}, nucleates and grows on silicate dust grains via surface hydrogenation reactions between atomic H and O \citep{greenberg1998,ruffle2001}. Its oxygen isotopic composition should therefore be similar to that of the aforementioned gaseous $^{18}$O-enriched atomic oxygen \citep{yurimoto2002}. The formation timescale for H$_2$O is about 10$^5$\,years \citep{bergin2000}. During this time, most of the atomic oxygen reacts to form H$_2$O ice, thus enriching the solid ice in $^{18}$O while simultaneously depleting the gas of $^{18}$O and also leaving CO as the most dominant gas species. In the case of CO$_2$, because it is produced via the reaction of CO with atomic O, its isotopic composition is between that of the $^{18}$O-enriched water and the $^{18}$O-depleted CO. \\

The isotopic fractionation of oxygen is subsequently \mbox{preserved} even if CO eventually becomes frozen onto the grains. This is because the isotopic exchange of oxygen between H$_2$O and CO ices is inefficient at low temperatures, according to \citet{yurimoto2004}. Their model further predicts that a direct measurement of cometary ices would yield a composition 5\% to 20\% enriched in $^{18}$O compared to terrestrial water. Our result, a $^{16}$O/$^{18}$O ratio of 445 $\pm$ 35 for cometary H$_2$O from 67P’s coma (an enrichment of 11\%), falls within this range and supports the prediction. \\

\newpage

\section{Conclusions}

From measurements of H${_2}^{16}$O / H${_2}^{18}$O and $^{16}$OH / $^{18}$OH \mbox{obtained} with the ROSINA DFMS on board the Rosetta \mbox{spacecraft}, and with our improved detector aging model, a $^{16}$O/$^{18}$O ratio of 445 $\pm$ 35 was found for H$_2$O in the coma of comet 67P. The evolution of our detector aging model to incorporate more sophisticated corrections to the gain of the detector is responsible for the differences between our result and an earlier report \citep{altwegg2014}. Our result, an \mbox{enrichment} of roughly 11\% of $^{18}$O as compared with the $^{16}$O/$^{18}$O ratio of 498.7~$\pm$~0.1 for terrestrial water \citep{baertschi1976}, is within the 5\% to 20\% range that leading self-shielding models \citep{sakamoto2007,yurimoto2004} predict for the \mbox{composition} of primordial water. \\

\vspace{1cm}

\begin{acknowledgements}

ROSINA would not have produced such outstanding results without the work of the many engineers, technicians, and scientists involved in the mission, in the Rosetta spacecraft, and in the ROSINA instrument team over the last 20 years, whose contributions are gratefully acknowledged. Rosetta is a European Space Agency (ESA) mission with contributions from its member states and NASA. We acknowledge herewith the work of the entire ESA Rosetta team. \\
\\
Work at the University of Bern was funded by the Canton of Bern, the Swiss National Science Foundation and the ESA PRODEX (PROgramme de D{\'e}veloppement d'Exp{\'e}riences scientifiques) programme. Work at the Southwest Research Institute was supported by subcontract {\#}1496541 from the Jet Propulsion Laboratory (JPL). Work at the Royal Belgian Institute for Space Aeronomy (BIRA-IASB) was supported by the Belgian Science Policy Office via PRODEX / ROSINA PRODEX Experiment Arrangement 90020. Work at the University of Michigan was funded by NASA under contract JPL-1266313. 

\end{acknowledgements}

\newpage

\bibliographystyle{aa}
\bibliography{references}

\begin{appendix}

\section{Supplementary material}

\begin{figure*}
\centering
        \includegraphics[width=17cm]{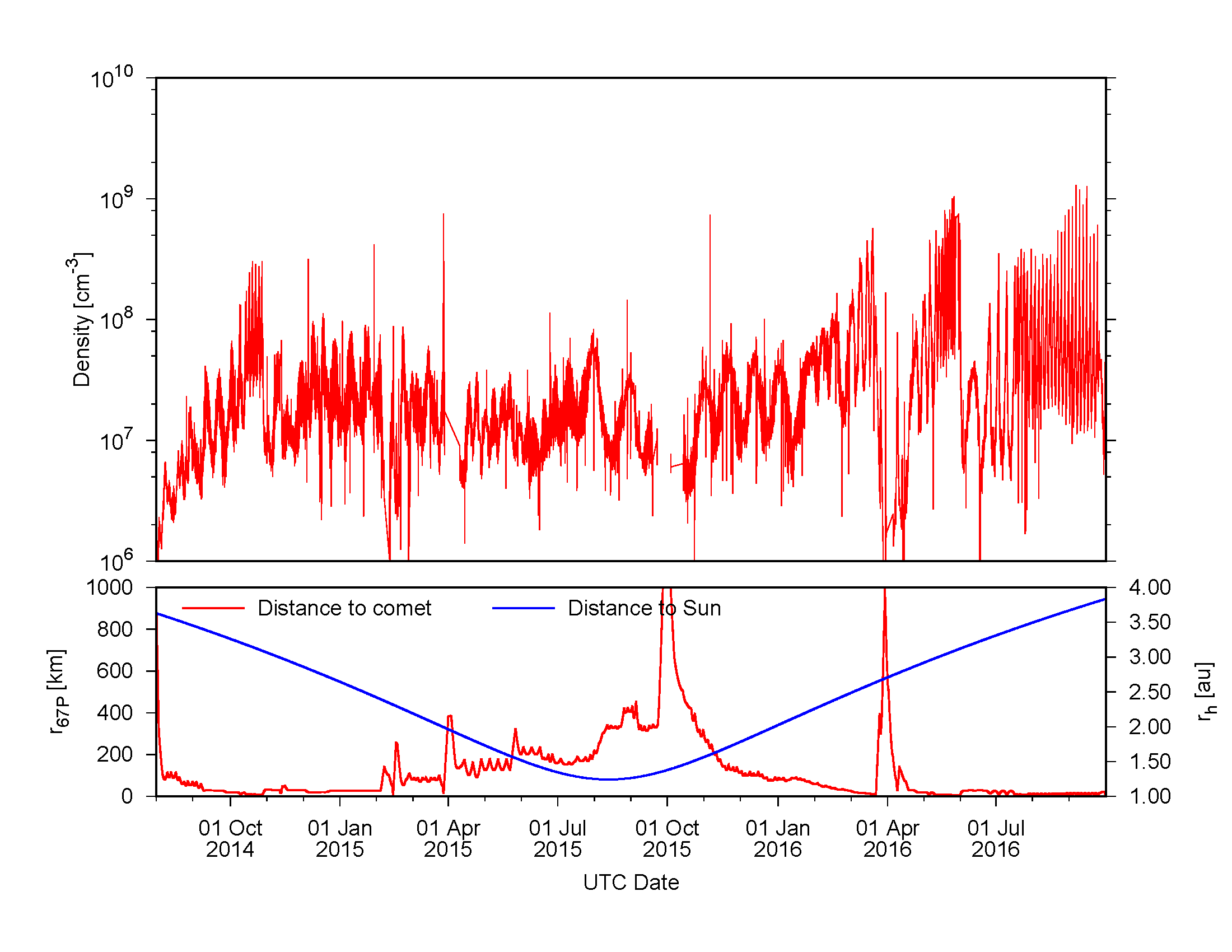}
        \caption{Particle density measured by COPS and heliocentric and cometocentric distance of the Rosetta spacecraft during its main mission duration.}
        \label{fig5}
\end{figure*}

The Cometary Pressure Sensor (COPS) on board the Rosetta spacecraft \citep{balsiger2007} measured the ambient particle density in the surrounding cometary coma that engulfed Rosetta as it accompanied comet 67P. The particle density measured by COPS, as well as the distance of Rosetta from comet 67P and its distance from the sun, are shown plotted over the main duration of the mission in \textbf{Fig. \ref{fig5}}. The figure shows that Rosetta generally remained within a few hundred kilometers of the comet after rendezvousing with it at around 3.5\,AU from the Sun, accompanying it to its perihelion at 1.24\,AU and then back out again to 3.5\,AU away from the Sun. The ambient particle density during this time was, according to COPS, typically in the range between 10$^7$ and 10$^8$ cm\textsuperscript{-3}. \\

\begin{figure}
        \resizebox{\hsize}{!}{\includegraphics{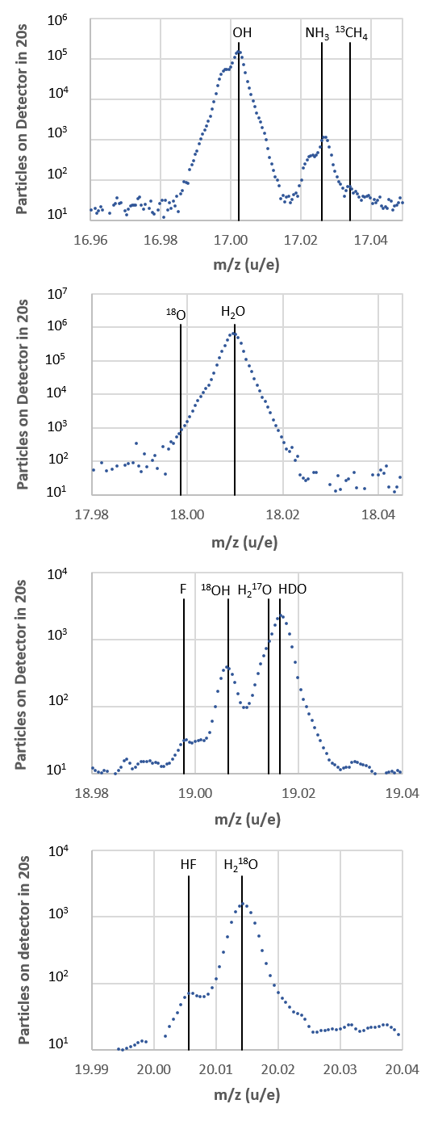}}
        \caption{Sample DFMS mass spectra (from 2014-10-20) for \mbox{m/z 17 to 20 u/e}.}
        \label{fig6}
\end{figure}

To give the reader an idea of what typical ROSINA DFMS mass spectra look like, a sample of DFMS mass spectra \mbox{measured} in 2014-10-20 is provided in \textbf{Fig. \ref{fig6}} for \mbox{mass-to-charge} ratios of 17 to 20 u/e. Minor deformation is seen in the shape of the peaks at m/z 17 u/e as a result of a slight instability in an electric potential in the electrostatic analyzer, the details and remedy for which were covered by \citet{dekeyser2015}. \\

\begin{figure*}
\centering
        \includegraphics[width=17cm]{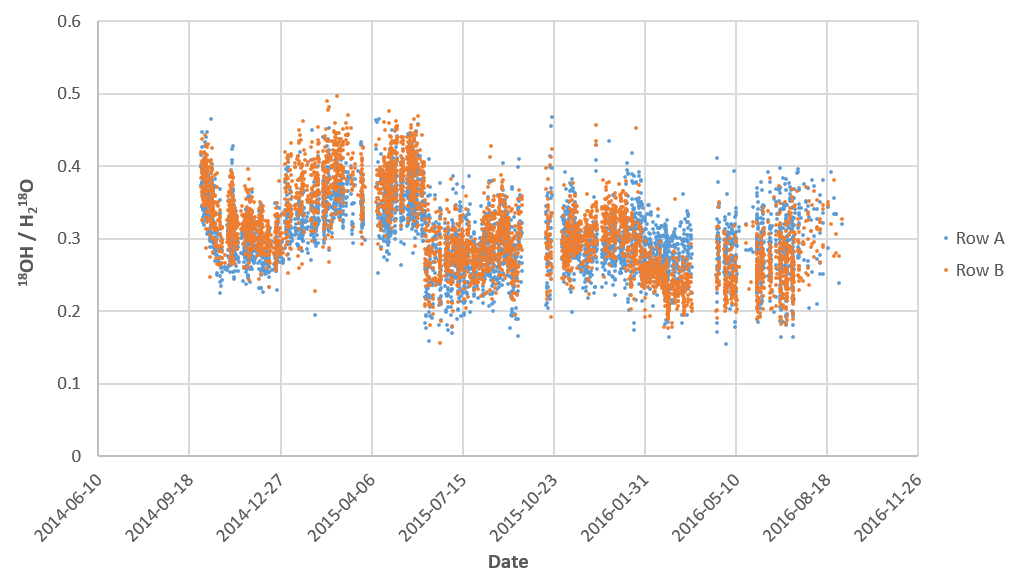}
        \caption{Uncorrected $^{18}$OH / H${_2}^{18}$O plotted against time of measurement.}
        \label{fig7}
\end{figure*}

As previously mentioned, several additional layers of corrections for changes in relative sensitivities, gain and pixel gain over time had to be applied to the DFMS data. To illustrate this, the uncorrected measurements of the fragmentation of water (the $^{18}$OH / H${_2}^{18}$O ratio) are shown plotted against the time at which they were measured in \textbf{Fig. \ref{fig7}}, using the original gain factors based on pre-flight calibrations. The fragmentation pattern of water should be constant, since the electron energy used by the DFMS for electron impact ionization was always 45 eV. It is clear from the figure, however, that there are sudden changes, the most abrupt being the one on 3$^{}$ June 2015, which was likely the result of damage. Although the nature and origin of the damage remain under debate, it is clear that it caused a change in instrument sensitivity. \\

\begin{figure*}
\centering
        \includegraphics[width=17cm]{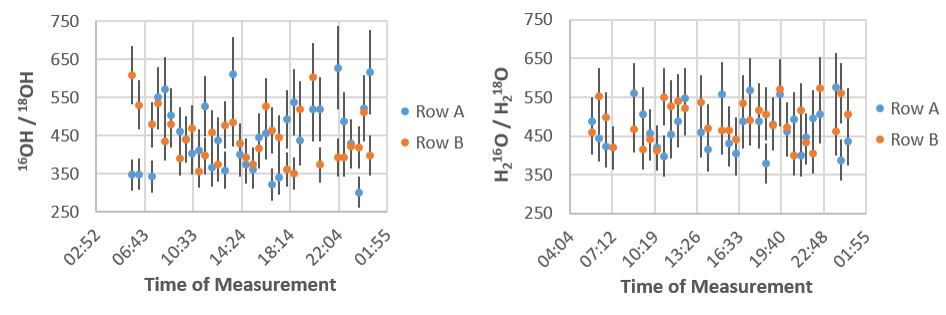}
        \caption{$^{16}$OH / $^{18}$OH (left) and H${_2}^{16}$O / H${_2}^{18}$O (right) measured during the sniff test on 19$^{}$ June 2014. Error bars reflect the uncertainties in the gain (6\%) and pixel gain (5\%) and the counting error.}
        \label{fig8}
\end{figure*}

Thus, to determine what the actual fragmentation pattern of water was, data from a sniff test conducted on 19$^{}$~June~2014 prior to Rosetta’s rendezvous with 67P had to be used. \mbox{Because} the water measured during this period was terrestrial \mbox{background} outgassed from Rosetta itself \citep{schlaeppi2010}, its isotopic composition, as previously demonstrated by \citet{haessig2013}, was the well-known terrestrial one, a fact that we exploited. The $^{16}$OH / $^{18}$OH and H${_2}^{16}$O / H${_2}^{18}$O ratios measured during the sniff test are shown in  \textbf{Fig. \ref{fig8}}. The mean $^{16}$OH~/~$^{18}$OH from the sniff test was  450 $\pm$ 22 and 445 $\pm$ 22 for Rows A and B, respectively, while the mean H${_2}^{16}$O / H${_2}^{18}$O was 470 $\pm$ 36 and 487 $\pm$ 38 for Rows A and B, respectively. By comparing this with the expected $^{16}$O/$^{18}$O of 498.7 $\pm$ 0.1 \citep{baertschi1976} of terrestrial water, we derived the following relative sensitivity corrections as given in \textbf{Table \ref{table1}}. \\

\begin{table}
\caption{Sensitivity relative to m/z 17 u/e.}
\label{table1}
\centering
\begin{tabular}{c c c c}  % no. of centered columns
\hline\hline  % inserts double horizontal lines
m/z (u/e) & species & Row A & Row B \\  % table headers
\hline  % inserts single horizontal line
        17 & $^{16}$OH          & 1                             & 1                     \\  % table contents
        18 & H${_2}^{16}$O      & 0.753 $\pm$ 0.070 & 0.761 $\pm$ 0.071 \\
        19 & $^{18}$OH          & 1.108 $\pm$ 0.055 & 1.119 $\pm$ 0.056 \\
        20 & H${_2}^{18}$O      & 0.799 $\pm$ 0.097 & 0.779 $\pm$ 0.094 \\
\hline  % inserts single line
\end{tabular}
\end{table}

The uncorrected mean $^{18}$OH / H${_2}^{18}$O from the sniff test was 0.461 $\pm$ 0.023 and 0.465 $\pm$ 0.023 for Rows A and B, respectively. We chose to use $^{18}$OH / H${_2}^{18}$O because, unlike $^{16}$OH and H${_2}^{16}$O, both $^{18}$OH and H${_2}^{18}$O were always measured on the same gain step (GS16) and their ratio would thus not be affected even if the gain (amplification) factors used were incorrect. Applying the relative sensitivity corrections from Table \ref{table1} to the fragmentation pattern from the sniff test yields a corrected $^{18}$OH / H${_2}^{18}$O of 0.33 $\pm$ 0.04. Having thus accurately determined the correct fragmentation pattern of water from the sniff test, we then proceeded to use it as a reference with which to derive corrections to account for subsequent changes in the relative sensitivities and gain that were caused by the aging of the detector. \\

By comparing the fragmentation of water in Fig. \ref{fig7} with the expected value of 0.33 $\pm$ 0.04, the following (multiplicative) correction factors, as presented in \textbf{Table \ref{table2}}, were derived to scale each approximately half-year interval to the correct value. \\

\begin{table}
\caption{Scaling factors for fragmentation of water.}
\label{table2}
\centering
\begin{tabular}{c c c}  % no. of centered columns
\hline\hline  % inserts double horizontal lines
interval & Row A & Row B \\  % table headers
\hline  % inserts single horizontal line
        date < 2014-12-29                                       & 1.076         & 0.994 \\  % table contents
        2014-12-29 $\leq$ date < 2015-06-03 & 0.957     & 0.849 \\
        2015-06-03 $\leq$ date < 2016-01-27 & 1.166     & 1.098 \\
        2016-01-27 $\leq$ date < 2016-04-26 & 1.198     & 1.332 \\
        2016-04-26 $\leq$ date                          & 1.159         & 1.237 \\
\hline  % inserts single line
\end{tabular}
\end{table}

Finally, to determine the gain factors for each period, the $^{16}$OH / H${_2}^{16}$O ratio was used. Unlike $^{18}$OH and H${_2}^{18}$O, $^{16}$OH and H${_2}^{16}$O were almost always measured on different gain steps and could therefore be used to compare the varying \mbox{differences} in the gain corresponding to different gain steps. \mbox{After} \mbox{application} of the scaling factors from Table \ref{table2} to \mbox{account} for sensitivity changes over time, any remaining deviation of $^{16}$OH~/~H${_2}^{16}$O from its expected value of 0.33 $\pm$ 0.04 would be due to changes in the gain. In this way, the gain factors, as presented in \textbf{Table~\ref{table3}}, could be derived from the fragmentation pattern of water. \\

\begin{table*}
\caption{Gain (i.e., amplification) factors. GS16 was kept fixed in the correction process.}
\label{table3}
\centering
\begin{tabular}{p{0.8cm} p{1.2cm} p{1.2cm} p{1.2cm} p{1.2cm} p{1.2cm} p{1.2cm} p{1.2cm} p{1.2cm} p{1.2cm} p{1.2cm}}  % no. of centered columns
\hline\hline  % inserts double horizontal lines
{} & \multicolumn{2}{p{2.4cm}}{\mbox{date <} \mbox{2014-12-29}} & \multicolumn{2}{p{2.4cm}}{\mbox{2014-12-29} \mbox{$\leq$ date <} \mbox{2015-06-03}} & \multicolumn{2}{p{2.4cm}}{\mbox{2015-06-03} \mbox{$\leq$ date <} \mbox{2016-01-27}} & \multicolumn{2}{p{2.4cm}}{\mbox{2016-01-27} \mbox{$\leq$ date <} \mbox{2016-04-26}} & \multicolumn{2}{p{2.4cm}}{2016-04-26 \mbox{$\leq$ date}} \\  % table headers
Gain Step & Row A & Row B & Row A & Row B & Row A & Row B & Row A & Row B & Row A & Row B \\
\hline  % inserts single horizontal line
        1       & 6.93          & 1.71          & 6.93          & 1.71          & 6.93            & 1.71          & 6.93          & 1.71          & 6.93          & 1.71            \\  % table contents
        2       & 9.48          & 3.43          & 9.48          & 3.43          & 9.48            & 3.43          & 9.48          & 3.43          & 9.48          & 3.43            \\
        3       & 14.6          & 7.23          & 14.6          & 7.23          & 14.6            & 7.23          & 14.6          & 7.23          & 14.6          & 7.23            \\
        4       & 24.9          & 15.6          & 24.9          & 15.6          & 24.9            & 15.6          & 24.9          & 15.6          & 24.9          & 15.6            \\
        5       & 48.5          & 36.7          & 48.5          & 36.7          & 48.5            & 36.7          & 48.5          & 36.7          & 48.5          & 36.7            \\
        6       & 107           & 93.5          & 107           & 93.5          & 107             & 93.5          & 107           & 93.5          & 107           & 93.5            \\
        7       & 248           & 238           & 248           & 238           & 248             & 238           & 248           & 238           & 248           & 238             \\
        8       & 640           & 652           & 640           & 652           & 640             & 652           & 640           & 652           & 640           & 652             \\
        9       & 1650          & 1730          & 1650          & 1730          & 1650            & 1730          & 1650          & 1730          & 1650          & 1730            \\
        10      & 6531          & 6877          & 4250          & 4480          & 5353            & 5007          & 5353          & 5007          & 5353          & 5007            \\
        11      & 13338         & 14732         & 19282         & 29026         & 14219           & 15005         & 13641         & 13404         & 15774         & 14090           \\
        12      & 26718         & 29130         & 40549         & 57124         & 38760           & 47575         & 28851         & 27257         & 33457         & 29165           \\
        13      & 76942         & 88234         & 104153        & 145105        & 108932  & 128053        & 83662         & 84038         & 95896         & 90543           \\
        14      & 197041        & 230712        & 246519        & 319780        & 237212  & 269549        & 206922        & 217531        & 225348        & 223725  \\
        15      & 471664        & 576457        & 554645        & 713801        & 582421  & 711066        & 615175        & 722174        & 647929        & 733282  \\
        16      & 1370000       & 1680000       & 1370000       & 1680000       & 1370000 & 1680000       & 1370000       & 1680000       & 1370000       & 1680000 \\
\hline  % inserts single line
\end{tabular}
\end{table*}

For our intended purposes, only the ratio between gain steps is important, since we used the DFMS to derive only the relative abundances of volatiles in the cometary coma. To derive total abundances, the relative abundances were then scaled such that the total particle density matched the total density measured by COPS, in the manner pioneered by \citet{gasc2017}. Thus, for the derivation of the gain factors shown in Table \ref{table3}, GS16 was chosen as the starting point and the corrections to the gain for the other gain steps were derived relative to GS16. Although several of the lower gain steps in certain intervals could not be corrected for changes over time due to a lack of data and were thus forced to retain their original gain factors, this is not an issue as they were seldom used, if ever. The evident changes in the gain reflect a decrease in the detector amplification of the higher gain steps with respect to the lower ones over time. \\

\begin{figure*}
\centering
        \includegraphics[width=17cm]{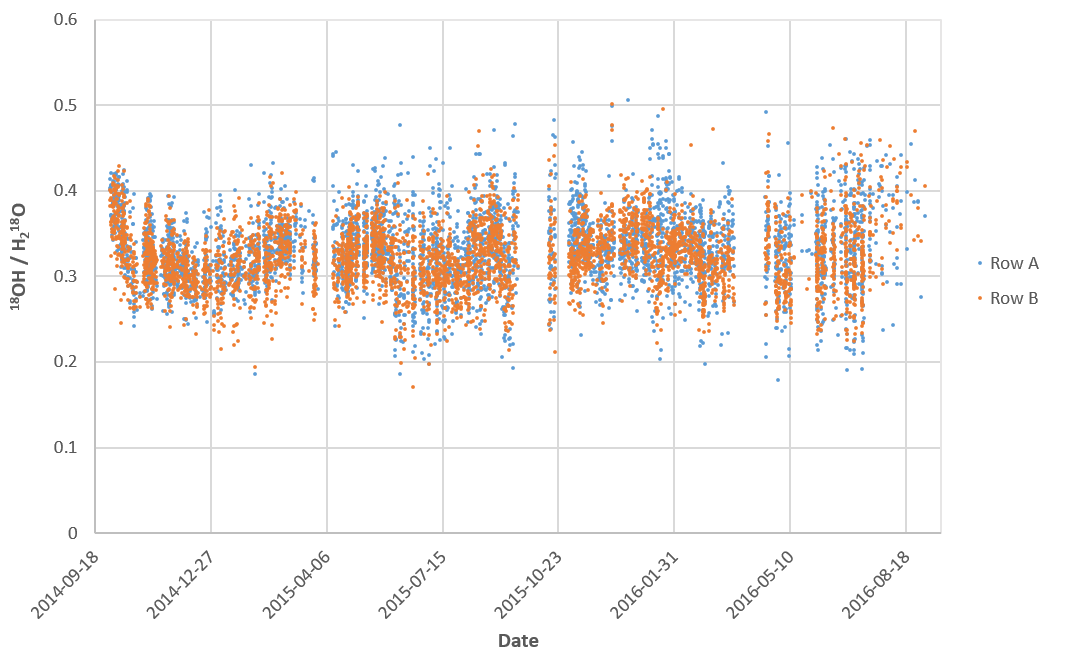}
        \caption{Corrected $^{18}$OH / H${_2}^{18}$O plotted against time of measurement.}
        \label{fig9}
\end{figure*}

Considering that both abrupt and gradual changes were observed over the course of the Rosetta space mission, the corrections in Table \ref{table3} had to be applied via a combination of step functions and linear interpolation over time. Step functions generally worked well for most of the intervals, with the exception of the period from 2016-01-27 to 2016-04-26. For this interval in particular, the median date (2016-03-12) represented the date when the gain factors were as displayed in Table \ref{table3} and the gain at any other time in this period was found by interpolation between this median date and the start or end of an adjacent interval. The result of this process is shown in \textbf{Fig.~\ref{fig9}}, where the corrected measurements of the fragmentation of water are plotted against the time at which they were measured. \\

\begin{figure*}
\centering
        \includegraphics[width=17cm]{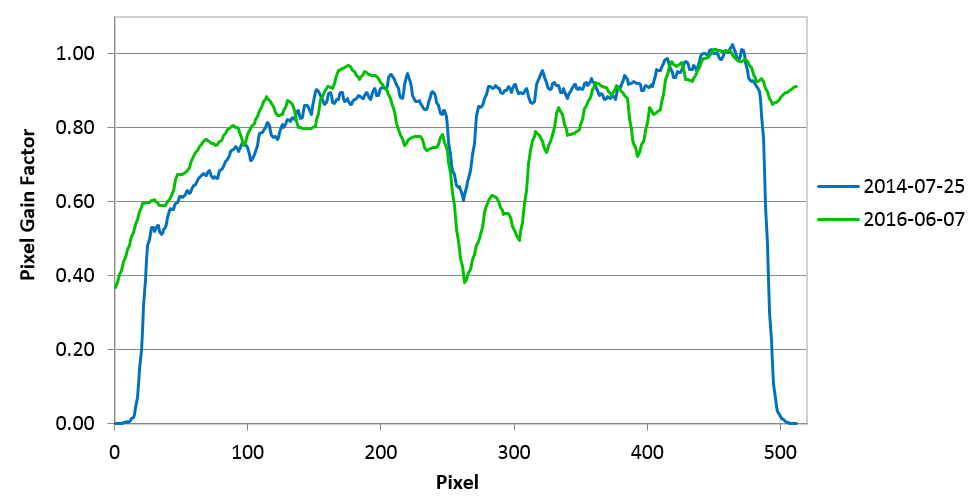}
        \caption{Pixel gain factors for each individual pixel on Row A for GS16. The 2014-07-25 and 2016-06-07 curves were measured early and late in the mission, respectively.}
        \label{fig10}
\end{figure*}

The final correction we shall mention, namely the pixel gain, was actually applied to the space data before the gain. We mention it last merely because it was directly measured at regular intervals throughout the mission. It accounts for the uneven degradation of individual pixels on the MCP / LEDA detector of the DFMS caused by uneven usage. To illustrate this, a sample of two pixel gain curves is shown in \textbf{Fig. \ref{fig10}}. One was measured early in the mission (2014-07-25) and the other near the end (2016-06-07). Both were measured for GS16 and show the pixel gain factors for each of the 512 individual pixels on Row A of the MCP / LEDA detector. Comparing the two curves, it is clear that the pixels in the center, which were used more frequently, were also the pixels that became the most heavily and quickly degraded over time as a result. \\

The DFMS had dedicated modes especially designed for the measurement of pixel gain, which were run during frequent campaigns conducted solely for that specific purpose. At times in between campaigns, linear interpolation over time was applied to derive appropriate pixel gain factors. \\

Our detector aging model incorporated all of these changes in relative sensitivity, gain, and pixel gain over time. The \mbox{accurate} determination of the isotopic composition of the cometary water of 67P subsequently depended upon the \mbox{application} of this model for the correction of measurements made with the DFMS. \\

\end{appendix}

\end{document}